\documentclass[12pt]{article}
\usepackage{amsfonts}
\usepackage{amssymb}
\usepackage{color}
\usepackage{graphicx}

\newcommand{\F}{{\cal F}}
\newcommand{\Sc}{{\cal S}}

\newcommand{\mc}{\mathcal}

\newcommand{\be}{\begin{equation}}
\newcommand{\en}{\end{equation}}
\newcommand{\bea}{\begin{eqnarray}}
\newcommand{\ena}{\end{eqnarray}}
\newcommand{\beano}{\begin{eqnarray*}}
\newcommand{\enano}{\end{eqnarray*}}

\newcommand{\V}{{\cal V}}

\newcommand{\ltwo}{{\Lc^2(\mathbb{R})}}

\newcommand{\EE}{\mc E}

\renewcommand{\l}{\langle}
\renewcommand{\r}{\rangle}
\newcommand{\pin}[2]{\l#1 , #2\r}

\newcommand{\Lc}{{\cal L}}
\newcommand{\1}{1 \!\! 1}

\newcommand{\Hil}{\mc H}
\newcommand{\Kc}{\mc K}

\catcode `\@=11 \@addtoreset{equation}{section}

\catcode `\@=12
\begin{document}

\begin{center}
{\Large \textbf{Ladder operators with no vacuum, their coherent states, and an application to graphene}} \vspace{2cm%
}\\[0pt]

{\large F. Bagarello}
\vspace{3mm}\\[0pt]
Dipartimento di Ingegneria,\\[0pt]
Universit\`{a} di Palermo, I - 90128 Palermo,\\
and I.N.F.N., Sezione di Catania\\
E-mail: fabio.bagarello@unipa.it\\

\vspace{7mm}

\end{center}

\vspace*{2cm}

\begin{abstract}
\noindent In literature ladder operators of different nature exist. The most famous are those obeying canonical (anti-) commutation relations, but they are not the only ones. In our knowledge, all ladder operators have  a common feature: the lowering operators annihilate a non zero vector, the {\em vacuum}. This is connected to the fact that operators of these kind are often used in factorizing some positive operators, or some operators which are { bounded from below}. This is the case, of course, of the harmonic oscillator, but not only. In this paper we discuss what happens when considering lowering operators with no vacua. In particular, after a general analysis of this situation, we propose a possible construction of coherent states, and we apply our construction to graphene. 
\end{abstract}

\vspace*{1cm}


\vfill

\newpage


\section{Introduction}\label{sect1}

The role of coherent states (CSs) in quantum mechanics is well recognized since its very beginning. Schr\"odinger himself introduced these vectors, \cite{schr}, as the {\em most classical} among all the quantum states. Since then, CSs have been used in quantum optics, quantization, quantum gravity, and in many other realms of quantum world. We refer to \cite{klauder}-\cite{ABG}, and references therein, just to have an idea of the huge class of applications of CSs considered along the years, and of the mathematical aspects which are somehow linked to their analysis.

One of the peculiarities of the CSs is that they are not really {\em uniquely defined}: different authors focus on different aspects of CSs, those which are more relevant for them, and because of these different points of view different expressions of CSs are sometimes proposed. In most cases, a coherent state is an eigenstate of some annihilation operator which resolves the identity and saturates the Heisenberg uncertainty principle, \cite{klauder}.

While in origin the annihilation operator was a bosonic operator, $a$, with $[a,a^\dagger]=\1$, it was soon realized that other operators also work in this analysis, and that different CSs can also be constructed, satisfying an eigenvalue equation with respect to other operators. Just to cite one class of CSs of this {\em extended} type, we cite here the so-called non linear CSs, \cite{manko,ali}, which are relevant for those Hamiltonian whose eigenvalues are not linear in their quantum number\footnote{As it happens for the harmonic oscillator.}. Another rather general class of generalized CSs are those known as {\em Gazeau-Klauder} CSs, \cite{GK}, where rather than focusing on ladder operators, one is directly interested in the Hamiltonian and in its eigenstates. In this case one of the crucial properties of CSs is that they are {\em temporally stable}: the time evolution of a coherent state is still coherent. The generalization to degenerate Hamiltonians was then proposed in \cite{bagali}. Quite recently, \cite{bagspringer}, another class of generalization of CSs has been proposed, the so-called bi-coherent states. These are pairs of eigenstates of two different annihilation operators, $A$ and $B^\dagger$, with $[A,B]=\1$ (in the sense of unbounded operators). These bi-coherent states also exist in a distributional settings, depending on the physical systems under consideration: these are calle {\em weak} bi-coherent states.

In all the situations listed above there is a common aspect: the (standard or generalized) CSs can be written as a series of vectors which includes a {\em vacuum}, i.e. a vector $e_0$ which is annihilated by the lowering operator of the physical system, $c$, $ce_0=0$, whichever its nature is, or which corresponds, see \cite{GK}, to the zero energy eigenvalue of an Hamiltonian. However, there also exist relevant physical systems which, in a natural way, give rise to sets of vectors which are eigenvectors of the Hamiltonian of the system, and which have no {\em natural vacuum}. Graphene is one of these systems. Another such system, a quantum particle on a circle, is considered in \cite{kow}. Graphene is our main motivation to check if and how a coherent state can be introduced out of an orthonormal (o.n.) set of vectors ${\varphi_p,\,p\in\mathbb{Z}}$, $\pin{\varphi_p}{\varphi_q}=\delta_{p,q}$, which is also total in a certain Hilbert space. This is the content of Section \ref{sect2}, while in Section \ref{sect3} we apply our general construction to graphene, and we discuss how to construct {\em ordinary} vector CSs for graphene. The construction we propose is, in our opinion, not trivial because of the use of unusual tools. Other attempts in constructing CSs for graphene are discussed, for instance, in \cite{fernandez,fernandez2}. Our approach is, we believe, more natural and the CSs we get are not particularly different from standard CSs. Our conclusions are given in Section \ref{sect4}.

\section{The abstract settings}\label{sect2}

Let $H$ be an operator acting on the Hilbert space $\Hil$, with scalar product $\pin{.}{.}$, linear in the second variable, and related norm $\|.\|=\sqrt{\pin{.}{.}}$. In what follows sometimes we will  call $H$ the {\em Hamiltonian} of a certain physical system $\Sc$, even if the eigenvalues of $H$ will not be assumed to be bounded from below (and from above). More explicitly, our working assumption is that we know the eigenvalues and the eigenvectors of $H$,
\be
H\varphi_p=\epsilon_p\varphi_p, 
\label{21}\en
where $p\in\mathbb{Z}$, with
\be
\pin{\varphi_p}{\varphi_q}=\delta_{p,q},
\label{22}\en
 $p, q\in\mathbb{Z}$, and $\cdots<\epsilon_{-2}<\epsilon_{-1}<\epsilon_{0}<\epsilon_{1}<\epsilon_{2}\cdots$. We are assuming therefore that the eigenvalues are real and strictly increasing, but not necessarily that $H=H^\dagger$. Moreover, it might happen that $\epsilon_p\rightarrow\pm\infty$ when $p\rightarrow\pm\infty$. In this case the eigenvalues of $H$ have no lower or upper bound, and it is not possible to consider any (finite) global shift of $H$, $H+\gamma\1$, for some real $\gamma$, so to get a new operator whose eigenvalues are bounded from below, or from above. Here $\1$ is the identity operator on $\Hil$. It is clear that, if the set $\{\epsilon_p\}$ has no lower or upper bounded, then $H$ is necessarily unbounded. In what follows we will also assume that the set $\F_\varphi=\{\varphi_p\}$ is total in $\Hil$: the only vector $f\in\Hil$ which is orthogonal to all the $\varphi_p$'s is the zero vector. In other words, $\F_\varphi$ is an o.n. basis for $\Hil$. We refer to \cite{kow} for an example of this settings, in the context of a particle moving on a circle.
 
 It is interesting to stress that we can always suppose that $\epsilon_p\neq0$, for all $p\in\mathbb{Z}$. Indeed, if for instance $\epsilon_0=0$, we can simply consider the new operator $\tilde H=H+\gamma\1$, where $\gamma=\frac{1}{2}\max\{\epsilon_{-1},\epsilon_1\}$. Then we have $\tilde H\varphi_p=\tilde\epsilon_p\varphi_p$, where $\tilde\epsilon_p=\epsilon_p+\gamma$, which is never zero: $\tilde H$ has the same eigenvectors as $H$, but its eigenvalues are all nonzero.
 
 \vspace{2mm}
 
 {\bf Remark:--} It is also interesting to see that it is not difficult to modify $H$ in order to obtain a different operator, with a finite number of eigenvalues which are different from those of $H$, while maintaining unchanged all its eigenvectors. This is, of course, different from what we did above, since in that case we have $\tilde\epsilon_p-\epsilon_p=\gamma$, $\forall p\in\mathbb{Z}$. Let now introduce a finite set of integers, $J=\{q_j\in\mathbb{Z}, \,j=1,2,\ldots,N\}$, $N<\infty$, and let $P_j$ be the orthogonal projector on $\varphi_{q_j}$: $P_jf=\pin{\varphi_{q_j}}{f}\varphi_{q_j}$. If we now {\em deform} $H$ as follows,
$$
H_J=H+\sum_{j=1}^N\delta_j\,P_j,
$$ 
$\delta_j\in\mathbb{R}$, $\delta_j\neq0$, we have the following: 
$$
H_J\varphi_p=\epsilon_p\varphi_p,
$$ 
 if $p\notin J$, while
 $$
 H_J\varphi_{q_j}=\hat\epsilon_{q_j}\varphi_{q_j},
 $$ 
if $q_j\in\ J$, where $\hat\epsilon_{q_j}=\epsilon_{q_j}+\delta_j$. In this way the eigenvectors of $H$ and $H_j$ coincide, while the eigenvalues of $H_J$ differ from those of $H$ only in a finite number, $N$. It is clear that, in this way, we can obtain from $H$ new operators with degenerate eigenvalues, but with eigenvectors which are  all mutually orthogonal.

 \vspace{2mm}
 
 Going back to our original Hamiltonian $H$, we will work under the assumption that, as already stated,  $\epsilon_p\neq0$, for all $p\in\mathbb{Z}$. This will be useful in the following. It is clear that it does not exist any operator $X$ on $\Hil$ such that $H=X^\dagger X$. This is not really related to the fact that $H$ is unbounded, even if in this situation domain issues may arise, but it is rather due to the fact that, even when $X^\dagger X$ is a well defined operator, $X^\dagger X$ must be positive, while $H$ is not. However, it is still open the possibility of factorizing $H$ as $H=ba$, for $a\neq b^\dagger$, since $ba$ is not necessarily positive. In fact, this is what we will do next. Let us now consider two sequences of (in general) complex numbers $\{\alpha_p, \,p\in\mathbb{Z}\}$, $\{\beta_p, \,p\in\mathbb{Z}\}$. We introduce two operators, $a$ and $b$, acting on $\F_\varphi$ as follows:
 \be
 a\varphi_p=\alpha_p\varphi_{p-1}, \qquad  b\varphi_p=\beta_{p+1}\varphi_{p+1},
 \label{23}\en
 where $p\in\mathbb{Z}$. The domains of these operators, $D(a)$ and $D(b)$, both contain $\Lc_\varphi=l.s.\{\varphi_p\}$, the linear span of the vectors $\varphi_p$. Of course, since $\F_\varphi$ is an o.n. basis for $\Hil$, $a$ and $b$ are densely defined. From (\ref{23}) we see that $a$ and $b$ are respectively a lowering and a raising operator. The main difference with respect to bosonic or pseudo-bosonic operators, see \cite{bagspringer,rom}, is that $a$ has no vacuum if  $\alpha_p\neq0$ for all $p\in\mathbb{Z}$. In particular, if $\alpha_0\neq0$, $a\varphi_0\neq0$. Similarly, if $\beta_p\neq0$ for all $p\in\mathbb{Z}$, $b$ has no vacuum as well, since $b\varphi_p\neq0$ for all $p\in\mathbb{Z}$. 
 
 It is easy to check that $a^\dagger$ and $b^\dagger$ act respectively as a raising and a lowering operator on $\F_\varphi$. Indeed we find that
 \be
 a^\dagger\varphi_p=\overline{\alpha_{p+1}}\varphi_{p+1}, \qquad  b^\dagger\varphi_p=\overline{\beta_{p}}\varphi_{p-1},
 \label{24}\en
 $\forall p\in\mathbb{Z}$. $a^\dagger$ and $b^\dagger$ are also densely defined. It is clear that $ a\varphi_p=\alpha_p\varphi_{p-1}=b^\dagger\varphi_p$ only if $\alpha_p=\beta_p$. However, this situation is not particularly interesting for us, since in this case we cannot use $a$ and $b$ to factorize $H$, as we have already observed. For this reason, in what follows we will always assume that $\alpha_p\neq\beta_p$, at least for some $p$. Using an useful bra-ket expression, we can rewrite
 \be
 a=\sum_{p\in\mathbb{Z}}\alpha_{p+1}|\varphi_p\rangle\langle \varphi_{p+1}|, \qquad b=\sum_{p\in\mathbb{Z}}\beta_{p+1}|\varphi_{p+1}\rangle\langle \varphi_{p}|,
 \label{25}\en
 with similar expansions for $a^\dagger$ and $b^\dagger$. Here we have $(|f\rangle\langle g|)h=\pin{g}{h}\,f$, $\forall f,g,h\in\Hil$. Using now (\ref{23}) it is clear that $ba\varphi_p=\alpha_p\beta_p\varphi_p$, and therefore
 \be
 H\varphi_p=ba\varphi_p \qquad \Leftrightarrow \qquad \alpha_p\beta_p=\epsilon_p,
 \label{26} \en
 $\forall p\in\mathbb{Z}$. In particular, because of our working assumption on the $\epsilon_p$'s, it follows that, $\forall p\in\mathbb{Z}$, $\alpha_p\neq0$ and $\beta_p\neq0$.
 
 The problem we would like to consider here is if it is possible to define a coherent state $\eta(z)$, depending on a complex variable $z$, such that $a\eta(z)=z\eta(z)$, for at least some $z\in\mathbb{C}$. However, the lack of a vacuum state in $\F_\varphi$ makes the existence of such a $\eta(z)$ impossible, at least using a standard approach, \cite{klauder}: we could, of course, look for an expansion of $\eta(z)$ in terms of $\varphi_p$, $\eta(z)\simeq \sum_{p\in\mathbb{Z}}k_pz^p\varphi_p$, for some (suitably chosen) complex sequence $\{k_p\}$. This series converge if $\sum_{p\in\mathbb{Z}}|k_p|^2|z|^{2p}<\infty$. But this is not a power series. This is a Laurent series which only converges for $z$ in some annulus in $\mathbb{C}$. In principle, this would not be a problem: literature on CSs contains several examples in which the states are defined not in all the complex plain, but only in some disk, \cite{aagbook,didier,gazeaubook,bagspringer}.  
 However, in our knowledge, there are no known examples of CSs in the literature in which the domain of convergence is an annulus. Moreover, and more essential, a state like $\eta(z)$ cannot be an eigenstate of $a$ in (\ref{23}), with eigenvalue $z$. This can be easily understood trying to extend the standard proof for CSs, \cite{klauder}, to the present settings. We see that this proof cannot work now, since $\{\varphi_p\}$ has no vector which is annihilated by $a$. For this reason, it is not so interesting to consider further $\eta(z)$ and its properties.  We will show that it is more efficient to slightly modify the Hilbert space we work with, and to introduce a new lowering operator in a {\em smart way}. This procedure will allow us to define {\em good} CSs.
 
 \subsection{From $\Hil$ to $\Hil_2$, and then to its restriction}\label{sect2.1}
 
 Let us now introduce the direct sum of $\Hil$ with itself, $\Hil_2=\Hil\oplus\Hil$:
 $$
 \Hil_2=\left\{f=\left(
 \begin{array}{c}
 	f_1 \\
 	f_2 \\
 \end{array}
 \right),\quad f_1,f_2\in\Hil
 \right\}.
 $$
 In $\Hil_2$ the scalar product $\left<.,.\right>_2$ is defined as
 \be
 \left<f,g\right>_2:=\left<f_1,g_1\right>+\left<f_2,g_2\right>,
 \label{27}\en
 and the square norm is $\|f\|_2^2=\|f_1\|^2+\|f_2\|^2$, for all $f=\left(
 \begin{array}{c}
 	f_1 \\
 	f_2 \\
 \end{array}
 \right)$, $g=\left(
 \begin{array}{c}
 	g_1 \\
 	g_2 \\
 \end{array}
 \right)$ in $\Hil_2$. In this space we introduce the following vectors:
 \be
 \Phi_p=\frac{1}{\sqrt{2}}\left(
 \begin{array}{c}
 	\varphi_p \\
 	\varphi_{-p} \\
 \end{array}
 \right), \qquad p\geq0
 \label{28}\en
 and the corresponding set $\F_\Phi=\{\Phi_p, p\geq0\}$. This set is o.n. in $\Hil_2$, $\pin{\Phi_p}{\Phi_q}_2=\delta_{p,q}$, but it is not total. In fact, for instance, the non zero vector $\left(
 \begin{array}{c}
 	\varphi_1 \\
 	-\varphi_{-1} \\
 \end{array}
 \right)$ is orthogonal to all the $\Phi_p$'s, $p\geq0$. Hence $\F_\Phi$ is not a basis for $\Hil_2$. But, if we consider the linear span of all the $\Phi_p$, $\Lc_\Phi=l.s.\{\Phi_p\}$, and we then take its completion in $\Hil_2$, we get a different Hilbert space $\Hil_\Phi=\overline{\Lc_\Phi}^{\|.\|_2}\subseteq\Hil_2$, and $\F_\Phi$ is o.n. and total in $\Hil_\Phi$. Therefore $\F_\Phi$ is an o.n. basis for $\Hil_\Phi$.  
 In what follows we will assume that 
 \be
 \alpha_p=\beta_{1-p}, \qquad \forall p\geq1.
 \label{29}\en
 This is not really a major constraint, in view of the great freedom we have on the sequences $\{\alpha_p\}$ and $\{\beta_p\}$. We will comment later on what can be done if (\ref{29}) is not satisfied. For the moment, we recall that $\{\alpha_p\}$ and $\{\beta_p\}$ are also assumed to satisfy (\ref{26}), to make of $H$ a factorizable operator. Let us now introduce the following orthogonal projectors, analogous to the $P_j$ we introduced before: $P_{\pm 1}f=\pin{\varphi_{\pm 1}}{f}\varphi_{\pm1}$, and then $Q_{\pm 1}=\1-P_{\pm 1}$. We have
 $$
 P_{\pm 1}^2= P_{\pm 1}^\dagger= P_{\pm 1}, \qquad Q_{\pm 1}^2= Q_{\pm 1}^\dagger= Q_{\pm 1},  
 $$
 and
 $$
P_{+1}P_{-1}=P_{-1}P_{+1}=0, \qquad Q_{+1}Q_{-1}=Q_{-1}Q_{+1}=\1-P_{-1}-P_{+1}.
 $$
 We use these operators to define the following new operator on $\Hil_2$:
 \be
 A=\left(
 \begin{array}{cc}
 	Q_{-1}a & 0 \\
 	0 & Q_1b \\
 \end{array}
 \right).
 \label{210}\en
 It is an easy exercise to check that $A$ behaves as a lowering operator on $\F_\Phi$. Indeed we have
 \be
 A\Phi_p=\theta_p\Phi_{p-1},\qquad p\geq0,
 \label{211}\en
 in which $\theta_0=0$, while $\theta_p=\alpha_p=\beta_{1-p}$ when $p\geq1$. From now on we will assume that each $\theta_p$ (and therefore $\alpha_p$ and $\beta_{1-p}$, $p\geq1$) is real. This is only meant to simplify the notation. The extension to complex $\theta_p$ is straightforward, but not useful for the application we will discuss in Section \ref{sect3}. The action of the adjoint of $A$ on each $\Phi_p$ is, as expected, $A^\dagger\Phi_p={\theta_{p+1}}\Phi_{p+1}$, $\forall p\geq0$, and therefore we have
 \be
 [A,A^\dagger]\Phi_p=\left(\theta_{p+1}^2-\theta_p^2\right)\Phi_p. \label{212}\en
 
 \vspace{2mm}
 
 {\bf Remarks:--} (1) It is maybe useful to explain why, in the definition (\ref{210}) of $A$ we use $a$ and $b$, rather than, say, $a$ and $a^\dagger$. This is indeed possible, in principle, since both $b$ and $a^\dagger$ act as raising operators on $\varphi_p$. However, as already observed, $a^\dagger$ cannot be used together with $a$ to factorize $H$. This is possible, see (\ref{26}), only when considering together $a$ and $b$. For this reason, we consider the one in (\ref{210}) the most natural definition of a lowering operator on $\F_\Phi$.
 
 (2) We recall that an operator analogous to our $A$ in (\ref{210}) has also been used in \cite{fernandez2}, in the attempt to introduce a lowering operator in connection with graphene. However, it should be stressed that the definition in \cite{fernandez2} is much more complicated than ours, and in this sense we believe that the one proposed here is a better. or at least a simpler, choice than that proposed in the cited paper.

 \vspace{2mm}

 Let us now define the function
 \be
 N(|z|)=\left(\sum_{k=0}^\infty\frac{|z|^{2k}}{(\theta_k!)^2}\right)^{-1/2},
 \label{213}\en
where we use the standard notation $\theta_0!=0!=1$, and $\theta_k!=\theta_1\theta_2\cdots\theta_k$, $k\geq1$. It is clear that $N(|z|)$ is well defined only for those $z\in\mathbb{C}$ for which the series $\sum_{k=0}^\infty\frac{|z|^{2k}}{(\theta_k!)^2}$ converges, and is different from zero. This is a power series, which converges for all $z\in C_\rho(0)$, the disk centered in the origin and with radius of convergence $\rho=\lim_{k,\infty}\theta_k$. For all these $z$ we can define the vector
\be
\Phi(z)=N(|z|)\sum_{k=0}^\infty \frac{z^k}{\theta_k!}\Phi_k.
\label{214}\en
Because of (\ref{211}), $\Phi(z)$ satisfies that following eigenvalue equation:
\be
A\Phi(z)=z\Phi(z), \qquad z\in C_\rho(0).
\label{215}\en
We stress once more that in this derivation we use, in particular, that $A\Phi_0=0$.
It is easy to find a condition which guarantees that $\Phi(z)$ produces a resolution of the identity: using the polar expression for $z$, $z=r\,e^{i\theta}$, let us consider a measure $d\nu(z,\overline z)=(N(|z|))^{-2}d\lambda(r)\,d\theta$. If $d\lambda(r)$ satisfies the following equality
\be
2\pi\int_{0}^\rho\,d\lambda(r)r^{2k}=(\theta_k!)^2,
\label{216}\en
for all $k\geq0$, then it is possible to check that
\be
\int_{C_\rho(0)}d\nu(z,\overline z)\pin{f}{\Phi(z)}_2\pin{\Phi(z)}{g}_2=\pin{f}{g}_2,
\label{217}\en
for all $f,g\in\Hil_\Phi$. We observe that $\Phi(z)$ does not resolve the identity in all of $\Hil_2$, and that the possibility solving the identity even only in $\Hil_\Phi$ is guaranteed only if we can find a solution of (\ref{216}), which is not always guaranteed, \cite{gazeaubook}. 

A standard coherent state $\phi(z)$ on $\ltwo$ is well known to saturate the Heisenberg uncertainty relation $\Delta x\,\Delta p=\frac{1}{2}$, where $\Delta S=\sqrt{\pin{\phi(z)}{S^2\phi(z)}-\pin{\phi(z)}{S\phi(z)}^2}=\sqrt{\langle S^2\rangle-\langle S\rangle^2}$, and where for instance $\langle S\rangle$ is the mean value of $S$ on $\phi(z)$. Here $S=x$ or $S=p$, the position and the momentum operators. These can be rewritten in terms of bosonic ladder operators $c$ and $c^\dagger$, $[c,c^\dagger]=\1$, as follows: $x=\frac{c+c^\dagger}{\sqrt{2}}$ and  $p=\frac{c-c^\dagger}{\sqrt{2}\,i}$. In this case we have $\phi(z)=e^{-|z|^2/2}\sum_{k=0}^\infty \frac{z^k}{\sqrt{k!}}\,e_k$, where each vector $e_k$ is constructed out of $c$ and $c^\dagger$: $ce_0=0$, and $e_k=\frac{(c^\dagger)^k}{\sqrt{k!}}\,e_0$, $k\geq1$. With this in mind, following the analogy with the above standard case, we introduce the following operators
$$
X=\frac{A+A^\dagger}{\sqrt{2}}, \qquad P=\frac{A-A^\dagger}{\sqrt{2}\,i},
$$
which are our counterparts of the position and momentum operators above, and we compute $\Delta X$ and $\Delta P$, replacing  $\varphi(z)$ with the vector $\Phi(z)$ in (\ref{214}). After some straightforward computations we find that
$$
(\Delta X)^2=(\Delta P)^2=\frac{1}{2}\left(\|A^\dagger\Phi\|_2^2-|z|^2\right),
$$
so that
\be
\Delta X\Delta P=\frac{1}{2}\left(\|A^\dagger\Phi\|_2^2-|z|^2\right),
\label{218}\en
It is interesting to notice that, if $A$ and $A^\dagger$ satisfy the canonical commutation relations  on $\Hil_\Phi$, $[A,A^\dagger]=\1_2$, i.e. if $\theta_p=\sqrt{p}$, $p\geq0$, then $\|A^\dagger\Phi\|_2^2=\pin{\Phi(z)}{AA^\dagger\Phi(z)}_2 =\pin{\Phi(z)}{(\1_2+A^\dagger A)\Phi(z)}_2 = 1+|z|^2 $, and therefore $\Delta X\Delta P=\frac{1}{2}$. In this case $\Phi(z)$ coincides with a two-components version of $\phi(z)$: $\Phi(z)$ is a vector coherent state, see \cite{alivcs1,alivcs2,bagali}. Notice that here we are using $\1_2$ to indicate the identity operator on $\Hil_2$.

\vspace{2mm}

{\bf Remark:--} As already mentioned, condition (\ref{29}) is useful, but not really essential. In fact, it would still be possible finding a lowering operator $\tilde A$ satisfying a lowering equality similar to the one in (\ref{211}). It is sufficient to replace $A$ in (\ref{210}) with
$$
 \tilde A=\left(
\begin{array}{cc}
	RQ_{-1}a & 0 \\
	0 & RQ_1b \\
\end{array}
\right),
$$
  where $R$ is a (densely defined) operator acting on $\varphi_p$ as follows: 
$$ R\varphi_p=\left\{
  \begin{array}{ll}
  	\alpha_{p+1}^{-1}\,\sqrt{p+1}\varphi_p, \hspace{5cm} p\geq0\\
  	\beta_{p}^{-1}\,\sqrt{1-p}\,\varphi_p, \hspace{5.1cm} p\leq0.\\
  \end{array}
  \right.
$$
  Of course, this definition makes sense, since all the $\alpha_p$ and $\beta_p$ are different from zero, but it imposes some constraint. In fact, $R$ is uniquely defined on $\varphi_0$ if $\alpha_1=\beta_0$, and this, together with (\ref{26}), introduces a new relation also between $\alpha_0$ and $\beta_1$. For instance, for graphene, this relation reads $3\alpha_0=\beta_1$. However, we will not insist on this alternative approach here, since the operator $A$ is already sufficient for our purposes.

\section{An application to graphene}\label{sect3}

The general construction outlined in Section \ref{sect2} is based on the existence of ladder operators for which no vacuum exists. This is exactly what happens when dealing with graphene, \cite{geim,nos1}. 
We devote the first part of this section to a very brief review of the construction of the eigenvectors of the Hamiltonian of the graphene in the Dirac points. In the second part we show how our previous analysis can be adapted to graphene, and in particular how CSs can be explicitly constructed.

\subsection{The eigenvectors of the graphene Hamiltonian}

We consider a layer of graphene, \cite{nos1} in an external constant magnetic field along $z$: $\vec B=B \hat e_3$. Adopting the symmetric gauge we have $\vec A=\frac{B}{2}(-y,x,0)$, and $\vec B=\nabla\wedge\vec A$. The Hamiltonian for the two Dirac points $K$ and $K'$ can be written as, \cite{geim},
\be
H_D=\left(
\begin{array}{cc}
	H_K & 0 \\
	0 & H_{K'} \\
\end{array}
\right),
\label{31}\en
where, in units $\hbar=c=1$, we have
\be
H_K=v_F\left(
\begin{array}{cc}
	0 & p_x-ip_y+\frac{eB}{2}(y+ix) \\
	p_x+ip_y+\frac{eB}{2}(y-ix) & 0 \\
\end{array}
\right).
\label{32}
\en
The operator $H_{K'}$ is just the transpose of $H_K$: $H_{K'}=H_K^T$, and $x,y,p_x$ and $p_y$ are the usual self-adjoint, two-dimensional position and momentum operators: $[x,p_x]=[y,p_y]=i\1$, all the other commutators being zero.  $\1$ is the identity operator in the relevant Hilbert space, which is now $\Kc:=\Lc^2(\Bbb R^2)$. The factor $v_F$ is the so-called Fermi velocity. The scalar product in $\Kc$ will be indicated as $\pin{.}{.}$.

Let us now introduce $\xi=\sqrt{\frac{2}{eB}}$, and the following canonical operators:
$$
X=\frac{1}{\xi}x,\qquad Y=\frac{1}{\xi}y,\qquad P_X=\xi p_x, \qquad P_Y=\xi p_y.
$$
These operators can be used to define two different pairs of bosonic operators: we first put $a_X=\frac{X+iP_X}{\sqrt{2}}$ and $a_Y=\frac{Y+iP_Y}{\sqrt{2}}$, and then
\be
A_1=\frac{a_X-ia_Y}{\sqrt{2}},\qquad A_2=\frac{a_X+ia_Y}{\sqrt{2}}.
\label{33}\en
The following commutation rules are satisfied:
\be
[a_X,a_X^\dagger]=[a_Y,a_Y^\dagger]=[A_1,A_1^\dagger]=[A_2,A_2^\dagger]=\1,
\label{34}
\en
the other commutators being zero. Then we have:
\be
H_K=\frac{2iv_F}{\xi}\left(
\begin{array}{cc}
	0 & A_2^\dagger \\
	-A_2 & 0 \\
\end{array}
\right),
\label{35}\en
which is Hermitian: $H_K=H_K^\dagger$. Since neither $H_K$ nor $H_{K'}$ depend on $A_1$ and $A_1^\dagger$, their eigenstates must be degenerate, \cite{bastos}. Let $e_{0,0}\in \Kc$ be the non zero vacuum of $A_1$ and $A_2$: $A_1e_{0,0}=A_2e_{0,0}=0$. Then we introduce, in standard fashion,
\be
e_{n_1,n_2}=\frac{1}{\sqrt{n_1!n_2!}}(A_1^\dagger)^{n_1}(A_2^\dagger)^{n_2}e_{0,0},
\label{36}
\en
and the set $\EE=\left\{e_{n_1,n_2},\, n_j\geq0\right\}$. $\EE$ is an o.n. basis for $\Kc$.   To deal with $H_K$ it is convenient to work in $\Kc_2=\Kc\oplus\Kc$, the direct sum of $\Kc$ with itself:
$$
\Kc_2=\left\{f=\left(
\begin{array}{c}
	f_1 \\
	f_2 \\
\end{array}
\right),\quad f_1,f_2\in\Kc
\right\}.
$$
In $\Kc_2$ the scalar product $\pin{.}{.}_2$ is defined as usual:
\be
\pin{f}{g}_2:=\pin{f_1}{g_1}+\pin{f_2}{g_2},
\label{37}\en
and the square norm is $\|f\|_2^2=\|f_1\|^2+\|f_2\|^2$, for all $f=\left(
\begin{array}{c}
	f_1 \\
	f_2 \\
\end{array}
\right)$, $g=\left(
\begin{array}{c}
	g_1 \\
	g_2 \\
\end{array}
\right)$ in $\Kc_2$. Introducing now the vectors
\be
e_{n_1,n_2}^{(1)}= \left(
\begin{array}{c}
	e_{n_1,n_2} \\
	0 \\
\end{array}
\right),\qquad e_{n_1,n_2}^{(2)}= \left(
\begin{array}{c}
	0 \\
	e_{n_1,n_2} \\
\end{array}
\right),
\label{38}\en
the set $\EE_2:=\{e_{n_1,n_2}^{(k)},\,n_1,n_2\geq0,\,k=1,2\}$  is an o.n. basis for $\Kc_2$. However, see \cite{bastos}, it is more convenient here to introduce the set $\V_2=\{v_{n_1,n_2}^{(k)},\,n_1,n_2\geq0,\,k=\pm\}$, where
\be
v_{n_1,0}^{(+)}=v_{n_1,0}^{(-)}=e_{n_1,0}^{(1)}=\left(
\begin{array}{c}
	e_{n_1,0} \\
	0 \\
\end{array}
\right),
\label{39}\en
while
\be
v_{n_1,n_2}^{(\pm)}=\frac{1}{\sqrt{2}}\left(
\begin{array}{c}
	e_{n_1,n_2} \\
	\mp i e_{n_1,n_2-1} \\
\end{array}
\right)=\frac{1}{\sqrt{2}}\left(e_{n_1,n_2}^{(1)}\mp i e_{n_1,n_2-1}^{(2)}\right),
\label{310}\en
if $n_2\geq1$. Quite often we will simply write $v_{n_1,0}=v_{n_1,0}^{(+)}=v_{n_1,0}^{(-)}$.
It is easy to check that these vectors are mutually orthogonal, normalized in $\Kc_2$, and total. Hence, $\V_2$ is an o.n. basis for $\Kc_2$. Its vectors are eigenvectors of $H_K$:
\be
H_K v_{n_1,0}=0, \quad H_K v_{n_1,n_2}^{(+)}=E_{n_1,n_2}^{(+)}  v_{n_1,n_2}^{(+)}, \quad   H_K v_{n_1,n_2}^{(-)}=E_{n_1,n_2}^{(-)}  v_{n_1,n_2}^{(-)},
\label{311}\en
where $E_{n_1,n_2}^{(\pm)}=\pm \frac{2v_F}{\xi}\,\sqrt{n_2}$. More compactly we can simply write $H_K v_{n_1,n_2}^{(\pm)}=E_{n_1,n_2}^{(\pm)} v_{n_1,n_2}^{(\pm)}$. We see explicitly that the eigenvalues have an infinite degeneracy in $n_1$, and that the set of the  $E_{n_1,n_2}^{(\pm)}$ is not bounded from below, nor from above.

Of course, both $\EE_2$ and $\V_2$  produce two different resolutions of the identity. Indeed we have
\be
\sum_{n_1,n_2=0}^\infty \sum_{k=1}^2\left<e_{n_1,n_2}^{(k)},f\right>_2 e_{n_1,n_2}^{(k)}=\sum_{n_1,n_2=0}^\infty \sum_{k=\pm}\left<v_{n_1,n_2}^{(k)},f\right>_2 v_{n_1,n_2}^{(k)}=f,
\label{312}\en
for all $f\in\Kc_2$.

Not many differences arise in the analysis of $H_{K'}$, since this is simply the transpose of $H_K$.

\subsection{Coherent states for $H_K$}

We will now discuss how, and in which sense, $H_K$ produces an explicit example of our general results in Section \ref{sect2}. However, since $E_{n_1,0}^{(\pm)}=0$, we need first of all to consider a proper shift of $H_k$, to get a different Hamiltonian whose eigenvalues are always different from zero. For that it is sufficient to define
\be
H=H_K+\frac{v_F}{\xi}\,\1_2, 
\label{313}\en
where $\1_2$, in this section, indicates the identity operator on $\Kc_2$. Hence we have 
$$
H v_{n_1,n_2}^{(\pm)}=\left(E_{n_1,n_2}^{(\pm)}+\frac{v_F}{\xi}\right) v_{n_1,n_2}^{(\pm)}=\frac{v_F}{\xi}\left(1\pm2\sqrt{n_2}\right) v_{n_1,n_2}^{(\pm)},
$$
for all $n_1,n_2=0,1,2,3,\ldots$. It is clear that none of the eigenvalues of $H$ is zero. It is also clear that, exactly as $E_{n_1,n_2}^{(\pm)}$, the set of all the eigenvalues of $H$ is unbounded, below and above. To adopt now what proposed in Section \ref{sect2}, we put
\be \varphi_p=\left\{
\begin{array}{ll}
	v_{n_1,p}^{(+)}, \hspace{5cm} p\geq1\\
	v_{n_1,0}^{(+)}=v_{n_1,0}^{(-)}=v_{n_1,0},  \hspace{2.4cm} p=0\\
	v_{n_1,-p}^{(-)}, \hspace{4.8cm} p\leq-1,\\
\end{array}
\right.
\label{314}\en
and
\be \epsilon_p=\left\{
\begin{array}{ll}
	\frac{v_F}{\xi}\left(1+2\sqrt{p}\right), \hspace{3.9cm} p\geq1\\
	\frac{v_F}{\xi},  \hspace{5.9cm} p=0\\
	\frac{v_F}{\xi}\left(1-2\sqrt{-p}\right), \hspace{3.6cm} p\leq-1.\\
\end{array}
\right.
\label{315}\en

We observe, first of all, that we are not making explicit the dependence on $n_1$: all the results which we will deduce from now on can be thought to be restricted to subspaces of Hilbert space corresponding to fixed values of $n_1$. This is very close to what is done when dealing with, say, the $n$-th Landau level for the Hall effect, \cite{chak}: each Landau level has an infinite degeneracy, and usually the interest is in what happens {\em inside} a single level, and in particular in the lowest Landau level, since this level corresponds to the minimum in energy of the physical system. 
With the above definitions we  are exactly in the conditions of Section \ref{sect2}. In particular we have $H\varphi_p=\epsilon_p\varphi_p$,
$
\pin{\varphi_p}{\varphi_q}_2=\delta_{p,q},
$
$p, q\in\mathbb{Z}$, and $\cdots<\epsilon_{-2}<\epsilon_{-1}<\epsilon_{0}<\epsilon_{1}<\epsilon_{2}\cdots$. Of course, $\epsilon_p\rightarrow\pm\infty$ when $p\rightarrow\pm\infty$. The only obvious difference is that the Hilbert space $\Hil$ in the abstract case is replaced here by $\Kc_2=\ltwo\oplus\ltwo$, with scalar product defined as in (\ref{37}). We can then proceed as in (\ref{23})-(\ref{25}) to define the ladder operators $a$ and $b$ acting on $\Kc_2$. For instance,
\be
af=\sum_{p=0}^\infty\alpha_{p+1}\pin{\varphi_{p+1}}{f}_2\,\varphi_p, \qquad b\,g=\sum_{p=0}^\infty\beta_{p+1}\pin{\varphi_{p}}{g}_2\,\varphi_{p+1},
\label{316}\en
for all $f\in D(a)$ and $g\in D(b)$. The sequences $\{\alpha_p\}$ and $\{\beta_p\}$ must be such that (\ref{26}) is satisfied, in order to factorize $H$. As already stressed, we assume that $\alpha_p$ and $\beta_p$ are real, for all $p\in\mathbb{Z}$. This is possible, since all the $\epsilon_p$ in (\ref{315}) are real. The sets $D(a)$ and $D(b)$, as already noticed, are dense in $\Kc_2$, since both contain the linear span of all the $\varphi_p$'s, $\Lc_\varphi$, which is dense in $\Kc_2$. Hence both $a$ and $b$ are densely defined. It is easy to compute the commutator between $a$ and $b$, and it is interesting to notice that this is independent of the particular choice of $\alpha_p$ and $\beta_p$, if (\ref{26}) is satisfied. We get
\be
[a,b]\varphi_p=\left(\epsilon_{p+1}-\epsilon_p\right)\varphi_p,
\label{add1}\en
$\forall p\in\mathbb{Z}$. Incidentally we observe that, because of (\ref{315}),
$$
\epsilon_{p+1}-\epsilon_p=\frac{2v_F}{\xi}\times\left\{
\begin{array}{ll}
\sqrt{p+1}-\sqrt{p}, \hspace{3.7cm} p\geq1\\
	1,  \hspace{5.9cm} p=-1,0\\
	\sqrt{-p}-\sqrt{-p-1}, \hspace{3.2cm} p\leq-2.\\
\end{array}
\right.
$$
We observe that it is exactly the presence of the square roots which makes of $[a,b]$ something different from the identity operator. Hence, in particular, $a$ and $b$ are not pseudo-bosonic operators, \cite{bagspringer}. 

The next step is to define CSs. However, as discussed in Section \ref{sect2.1}, it is convenient to double the Hilbert space we work with. In other words, we need to move from $\Kc_2$ to $\Kc_4=\Kc_2\oplus\Kc_2$. We call $\pin{.}{.}_4$ its scalar product, and $\|.\|_4$ its related norm. Of course, these are defined as in Section \ref{sect2.1}:

$$
\Kc_4=\left\{f=\left(
\begin{array}{c}
	f_1 \\
	f_2 \\
\end{array}
\right),\quad f_1,f_2\in\Kc_2
\right\},
$$
with
$$
\left<f,g\right>_4:=\left<f_1,g_1\right>_2+\left<f_2,g_2\right>_2,
$$
and  $\|f\|_4^2=\|f_1\|_2^2+\|f_2\|_2^2$, for all $f=\left(
\begin{array}{c}
	f_1 \\
	f_2 \\
\end{array}
\right)$, $g=\left(
\begin{array}{c}
	g_1 \\
	g_2 \\
\end{array}
\right)$ in $\Kc_4$. In analogy with what seen before we introduce the following vectors:
\be
\Phi_p=\frac{1}{\sqrt{2}}\left(
\begin{array}{c}
	\varphi_p \\
	\varphi_{-p} \\
\end{array}
\right), \qquad p\geq0,
\label{317}\en
which are now clearly vectors in $\Kc_4$,
and the corresponding set $\F_\Phi=\{\Phi_p, p\geq0\}$, which is o.n. in $\Kc_4$, $\pin{\Phi_p}{\Phi_q}_4=\delta_{p,q}$, but not total, as shown before. For this reason we consider (we use the same notation as in Section \ref{sect2.1}, here) the linear span of all the $\Phi_p$, $\Lc_\Phi=l.s.\{\Phi_p\}$, and we then take its completion in $\Kc_4$, defining a different Hilbert space $\Kc_\Phi=\overline{\Lc_\Phi}^{\|.\|_4}\subseteq\Kc_4$. $\F_\Phi$ is o.n. and total in $\Kc_\Phi$, and therefore it is a basis in $\Kc_\Phi$. Our main interest here is the construction of coherent states. We remind the reader that, in (\ref{316}), we have $\alpha_p=\beta_{1-p}$, $\forall p\geq1$.

Repeating now the general construction proposed in Section \ref{sect2.1}, we introduce next the orthogonal projectors $P_{\pm1}$ and $Q_{\pm1}=\1_2-P_{\pm1}$, where $\1_2$ is the identity operator on $\Kc_2$. Then we define an operator $A_4$ as in (\ref{210}):
\be
A_4=\left(
\begin{array}{cc}
	Q_{-1}a & 0 \\
	0 & Q_1b \\
\end{array}
\right).
\label{318}\en
Despite of the fact that $A$ and $A_4$ look exactly the same, ot should be stressed that $A$ acts on the generic Hilbert space $\Hil_2$, while $A_4$ acts on the concrete Hilbert space $\Kc_4$ introduced ad hoc to deal for the Hamiltonian $H_D$ in (\ref{31}).
We have 
\be
A_4\Phi_p=\theta_p\Phi_{p-1},\qquad p\geq0,
\label{319}\en
where, as in Section \ref{sect2.1}, $\theta_0=0$, while $\theta_p=\alpha_p=\beta_{1-p}$ when $p\geq1$. Hence $A_4^\dagger\Phi_p=\theta_{p+1}\Phi_{p+1}$, $p\geq0$. Apart from our working condition $\alpha_p=\beta_{1-p}$, $\forall p\geq1$, we recall that we are interested in having $\alpha_p\beta_p=\epsilon_p$, where $\epsilon_p$ are those in (\ref{315}). Of course, there are many possible choices of $\{\alpha_p\}$ and $\{\beta_p\}$ which obey these conditions. We consider here few of them. In particular, and this will be our Choice 3 below,
 we show that it is possible to operate a special choice which produces a sort of standard CSs, satisfying on $\Kc_\Phi$ all the properties of ordinary CSs, \cite{klauder}.

We start considering the following choice:

\vspace{2mm}

 {\bf Choice 1:--}
\be \alpha_p=\left\{
\begin{array}{ll}
	1, \hspace{3.1cm} p\geq1\\
	\frac{v_F}{\xi}\left(1-2\sqrt{-p}\right), \hspace{0.6cm} p\leq0,
\end{array}
\right.\qquad \beta_p=\left\{
\begin{array}{ll}
	\frac{v_F}{\xi}\left(1+2\sqrt{p}\right), \hspace{0.6cm} p\geq1,\\
	1, \hspace{2.8cm} p\leq0.
\end{array}
\right.
\label{320}\en
We see that $\{\alpha_p\}$ is bounded from above, but not from below. The opposite is true for $\{\beta_p\}$. With this choice it is quite easy to compute most of the  quantities which are relevant for us. In particular, since $\theta_0=0$, and $\theta_p=1$ for all $p\geq1$, using (\ref{212}) we have that $[A_4,A_4^\dagger]=|\Phi_0\rangle\langle\Phi_0|$, at least on $\Lc_\Phi$. Moreover, $\theta_p!=1$ for all $p\geq0$. Hence (\ref{213}) and (\ref{214}) take the following forms:

$$
N(|z|)=\left(\sum_{k=0}^\infty|z|^{2k}\right)^{-1/2}=\sqrt{1-|z|^2},
$$
 for all $z\in C_1(0)$, the disk $|z|<1$ in the complex plane. For all these $z$ we can define the vector
\be
\Phi(z)=N(|z|)\sum_{k=0}^\infty z^k\Phi_k,
\label{321}\en
and, because of (\ref{319}), $\Phi(z)$ satisfies that following eigenvalue equation:
\be
A_4\Phi(z)=z\Phi(z), \qquad z\in C_\rho(0).
\label{322}\en
As we see from (\ref{218}), to compute $\Delta X\,\Delta P$, we need to compute $\|A_4^\dagger\Phi\|_4$ first. In our case, the computation is simple:
$$
\|A_4^\dagger\Phi\|_4^2=\pin{A_4^\dagger\Phi(z)}{A_4^\dagger\Phi(z)}_4=\pin{\Phi(z)}{[A_4,A_4^\dagger]\Phi(z)}_4+\pin{\Phi(z)}{A_4^\dagger A_4\Phi(z)}_4=
$$
$$
|\pin{\Psi_0}{\Psi(z)}|^2+\|A_4\Phi(z)\|^2=(\sqrt{1-|z|^2})^2+|z|^2=1.
$$
Hence we have
\be
\Delta X\,\Delta P=\frac{1}{2}\left(1-|z|^2\right).
\label{323}\en
Since, on $\Lc_\Phi$, $[X,P]=\frac{1}{2i}[A_4+A_4^\dagger,A_4-A_4^\dagger]=i[A_4,A_4^\dagger]=i|\Phi_0\rangle\langle\Phi_0|$, we easily conclude that $\Delta X\,\Delta P=\frac{1}{2}|\pin{\Phi(z)}{[X,P]\Phi(z)}|$: the vector $\Phi(z)$ saturates the Heisenberg uncertainty relation. This is indeed what one would like to obtain from $\Phi(z)$ to start considering it a {\em coherent state}. However, we would also expect that $\Phi(z)$ solves the identity as in (\ref{217}). However, in this case condition (\ref{216}) becomes
$
2\pi\int_{0}^1\,d\lambda(r)r^{2k}=1$ for all $k\geq0$, which has no (non-distributional) solution. Hence the (partial) conclusion is that the state (\ref{321}) does not resolve the identity. So the question is: can we do better? In other words, can we operate a different choice of $\{\alpha_p\}$ and $\{\beta_p\}$ such that another vector $\Phi(z)$ can be introduced \underline{also} resolving the identity?

 The answer is affirmative. However, before we discuss how this can be done, it might be interesting to observe that choice (\ref{320}) is not the worst choice we could consider, at least from a practical point of view. Let us indeed consider the following alternative, which is, in a sense, specular to Choice 1\footnote{We mean that, while in Choice 1 we had $\alpha_p=1$ for all $p\geq1$, now the same requirement is asked to $\beta_p$, $p\geq1$.}:

\vspace{2mm}

 {\bf Choice 2:--}
\be \alpha_p=\left\{
\begin{array}{ll}
	\frac{v_F}{\xi}\left(1+2\sqrt{p}\right), \hspace{0.6cm} p\geq1,\\
	\frac{1-2\sqrt{-p}}{1+2\sqrt{1-p}}, \hspace{16mm} p\leq0
\end{array}
\right.\qquad \beta_p=\left\{
\begin{array}{ll}
	1, \hspace{3.5cm} p\geq1,\\
	\frac{v_F}{\xi}\left(1+2\sqrt{1-p}\right), \hspace{0.6cm} p\leq0.
\end{array}
\right.
\label{324}\en
This is another solution of our conditions $\alpha_p=\beta_{1-p}$, $\forall p\geq1$ and $\alpha_p\beta_p=\epsilon_p$, $p\in\mathbb{Z}$. Then $\theta_0=0$, while $\theta_p=\alpha_p=\frac{v_F}{\xi}\left(1+2\sqrt{p}\right)$, $p\geq1$. It is clear that $\rho=\infty$, so that the state $\Phi(z)$ in (\ref{214}) exists for all $z\in\mathbb{C}$. However, it is very complicated (and not particular interesting) to check if $\Phi(z)$ solves the identity (and finding the relevant measure) or if it saturates the uncertainty inequality. So the choice in (\ref{324}) is possible, but is technically very complicated. It is more convenient, and much easier, to consider the alternative below.  

\vspace{2mm}

 {\bf Choice 3:--}
\be \alpha_p=\left\{
\begin{array}{ll}
	\sqrt{p}, \hspace{3.2cm} p\geq1\\
	\frac{v_F}{\xi\sqrt{1-p}}\left(1-2\sqrt{-p}\right), \hspace{0.5cm} p\leq0,
\end{array}
\right.\qquad \beta_p=\left\{
\begin{array}{ll}
	\frac{v_F}{\xi\sqrt{p}}\left(1+2\sqrt{p}\right), \hspace{0.6cm} p\geq1,\\
	\sqrt{1-p}, \hspace{2.cm} p\leq0.
\end{array}
\right.
\label{325}\en
With this choice, $\theta_p=\sqrt{p}$, for all $p\geq0$, so that $A$ and $A^\dagger$ satisfy, on $\Lc_\Phi$, the CCR: $[A_4,A_4^\dagger]f=f$, $\forall f\in\Kc_4$. Hence $\Phi(z)$ is a standard coherent state and, as such, satisfies all the properties of this class of states: $\Phi(z)$ is an eigenstate of the lowering operator $A_4$, saturates the uncertainty inequality, and resolves the identity: in this case the measure $d\nu(z,\overline z)$ in (\ref{217}) is easy:  $d\nu(z,\overline z)=\frac{1}{\pi}\,r\,dr\,d\theta$. Then the conclusion is that, even for the Hamiltonian $H_K$ considered here, we can introduce CSs as those arising in the analysis of the harmonic oscillator, for an operator $A_4$ which is not, of course, the standard lowering operator of an harmonic oscillator, even if it still satisfies the canonical commutation relations. This choice is particularly useful, since it allows to factorize $H$ and to define easily CSs with all the best properties we could desire, contrarily to what happens for the Choices 1 and 2.

\section{Conclusions}\label{sect4}

In this paper we have discussed some aspects of ladder operators defined on a set on infinite vectors without a {\em vacuum}, i.e. a set in which no vector is annihilated by any of the ladder operators, or their adjoints. These operators turn out to be useful to factorize Hamiltonians which are not bounded below or above. This is the case of the Hamiltonian for the graphene in the Dirac points. We use our construction to define possible CSs in a general settings, and for graphene, and we show that it is indeed possible to define vectors which have all the properties of standard CSs, at the (small) price of working in larger Hilbert spaces, and to define properly the ladder operators which are used to factorize the Hamiltonian.

\section*{Acknowledgements}

The author is indebted to J.P. Gazeau for an useful suggestion on the solution of the moment problem in Section \ref{sect2.1}.

\section*{Data accessibility statement}

This work does not have any experimental data.

%
%
%
%
%

\section*{Funding statement}

The author acknowledges partial financial support from Palermo University, via FFR2021 "Bagarello", and from G.N.F.M. of the INdAM. This work has also been partially supported by the PRIN grant {\em Transport phenomena in low dimensional
structures: models, simulations and theoretical aspects}.


\begin{thebibliography}{99}
	
	\bibitem{schr} E. Schr\"odinger,  {\em Der stetige \"Ubergang von der Mikro- zur Makromechanik}, Die Naturwissenschaften,  {\bf 14}, 664–666 (1926), translated in {\em  The Continuous Transition from Micro- to
	Macro-Mechanics}, in E. Schr\"odinger, {\em Collected Papers in Wave Mechanics}, London: Blackie$\&$Son,
	41-44 (1928)
	
	\bibitem{klauder} J. R. Klauder, B. S. Skagerstam Eds., {\em Coherent states. Applications in physics and mathematical physics}, World Scientific, Singapore (1985)
	
			\bibitem{aagbook}  S.T. Ali, J-P.  Antoine and  J-P.  Gazeau,
	{\em  Coherent States, Wavelets and Their Generalizations\/},
	Springer-Verlag, New York, (2000).
	
	\bibitem{didier} M. Combescure,  R. Didier, {\em Coherent States and Applications in Mathematical Physics},   Springer, (2012)
	
	\bibitem{gazeaubook}  J-P.  Gazeau, {\em Coherent states in quantum physics}, Wiley-VCH, Berlin (2009)
	
	\bibitem{perelomov} A. M. Perelomov, {\em Generalized coherent states and their applications}, Springer-Verlag, Berlin (1986)
	
	\bibitem{BAAG}  S. T. Ali, J.-P. Antoine, F. Bagarello, J.-P. Gazeau, Guest Editors, {\em Coherent states: mathematical
		and physical aspects}, Journal of Physics A: Mathematical and Theoretical, Special Issue, {\bf 45}, N. 24, (2012)
	
	\bibitem{ABG}   J.-P. Antoine, F. Bagarello, J.-P. Gazeau Eds, {\em Coherent states	and applications: a contemporary panorama}, Springer Proceedings in Physics, (2018)
	
	
	\bibitem{manko} M. V. Manko, G. Marmo, E. C. G. Sudarshan, F. Zaccaria, {\em $f$-oscillators and nonlinear coherent states}, Physica Scripta. {\bf 55} (5): 528–541 (1997)
	
	\bibitem{ali} S.T. Ali, Z. Mouayn, K. Ahbli, {\em Nonlinear Coherent States Associated with a Measure on the Positive Real Half Line}, Complex Anal. Oper. Theory {\bf 14}, 24 (2020)

	\bibitem{GK} J.P. Gazeau, J.R. Klauder, {\em Coherent states for systems with discrete and continuous
		spectrum}, J. Phys. A, {\bf 32}, 123-132 (1999)
	
	\bibitem{bagali} F. Bagarello, S.T. Ali, {\em Some Physical Appearances of
		Vector Coherent States and Coherent States Related to Degenerate Hamiltonians},
	J. Math. Phys, {\bf 46},
	053518 (2005)
	


\bibitem{bagspringer} F. Bagarello, {\em Pseudo-bosons and their coherent states}, Springer (2022)

	\bibitem{kow} K. Kowalskiy, J. Rembielinski, L. C. Papaloucas, {\em Coherent states for a quantum particle on a circle}, J. Phys. A, {\bf 29}, 4149-4167 (1996)
	
		\bibitem{fernandez} D. J. Fernandez C, D. O-Campa, {\em Graphene generalized coherent states}, Eur. Phys. J. Plus, {\bf 137}, 1012 (2022).
	
	
	\bibitem{fernandez2} E. Diaz-Bautista, D. J. Fernandez C,  {\em Graphene coherent states}, Eur. Phys. J. Plus, {\bf 132}, 499 (2017).
	
	
	
	
	
	
	\bibitem{rom} P. Roman, {\em Advanced quantum mechanics}, Addison--Wesley, New York,
	1965.
	
	
	
	\bibitem{alivcs1}  K. Thirulogasanthar and S.T. Ali, {\em A class of vector coherent states defined
		over matrix domains}, J. Math. Phys. {\bf 44},  5070-5083 (2003)
	
	\bibitem{alivcs2} S.T. Ali, M. Engli\v s and J.-P. Gazeau, {\em Vector coherent states from
		Plancherel's theorem, Clifford algebras and matrix domains}, J. Phys. {\bf A37},
	 6067-6089 (2004)
	
	
	

	
		\bibitem{geim} A.K. Geim, K.S. Noselov, {\em The rise of Graphene}, Nature Materials, {\bf 6}, (3), 183-191 (2007)
	
	
	\bibitem{nos1} A. K. Geim, {\em Graphene: Status and Prospects}, Science, {\bf 324}, 1530-1534, (2009); A. H. Castro Neto, F. Guinea, N. M. R. Peres, K. S. Novoselov, and A. K. Geim, {\em The electronic properties of graphene}, Rev. Mod. Phys., {\bf 81}, 109-162 (2009); V. Chabot,  D. Higgins, A. Yu,   X. Xiao,   Z. Chena and    J. Zhang,  {\em A review of graphene and graphene oxide sponge: material synthesis and applications to energy and the environment}, Energy Environ. Sci., {\bf 7}, 1564-1596 (2014)
	
	
	\bibitem{bastos} C. Bastos, O. Bertolami, N. Costa Dias, J. Nuno Prata, {\em Noncommutative Graphene}, Int. J. Mod. Phys. A {\bf 28},  1350064 (2013)
	
	
	\bibitem{chak} T. Chakraborty, P. Pietil\"ainen, {\em The fractional quantum Hall effect-Properties of an incompressible quantum fluid}, Springer Berlin, Heidelberg (2012)   
	
	
	
	
	
	
	
	
	
	
	
	
	
	
	
	
	
	
	
	
	
	
	
	
	



	


\end{thebibliography}
\end{document}